\documentclass[aps,pre,preprint,groupedaddress]{revtex4-1}
\usepackage{graphicx}% Include figure files

\begin{document}

%\preprint{}
%%\begin{CJK*}{UTF8}{}

\title{Crumpling transition of the
discrete planar folding in the negative-bending-rigidity regime} 

\author{Yoshihiro Nishiyama} %% (西山由弘)  }
%\thanks{}
\affiliation{Department of Physics, Faculty of Science,
Okayama University, Okayama 700-8530, Japan}

\date{\today}

\begin{abstract}
The folding of the triangular lattice
embedded in two dimensions
(discrete planar folding)
is investigated numerically.
As the bending rigidity $K$ varies,
the planar folding exhibits
a series of crumpling transitions
at $K \approx -0.3 $ and $K \approx 0.1$.
By means of the transfer-matrix method for the system sizes 
$L \le 14$,
we analyze the singularity of the transition at $K \approx -0.3$.
As a result,
we estimate the transition point
and the latent heat
as 
$K=-0.270(2)$ and 
$Q=0.043(10)$, respectively.
This result suggests
that the singularity 
belongs to
a weak-first-order transition.
\end{abstract}

%\keywords{}

\maketitle

%%\end{CJK*}

%%%\section{\label{section1}
%%%Introduction}
%\subsection{}
%\subsubsection{}

At sufficiently low temperatures,
a polymerized membrane 
becomes flattened macroscopically
\cite{Nelson87}; see Refs.
\cite{Nelson89,Nelson96,Bowick01,Gompper97}
for a review.
%The flat phase is characterized by
%the long-range order of the surface normals.
%[It is rather exceptional that 
%an orientational (continuous) symmetry
%breaks 
%spontaneously for such a two-dimensional manifold.]
It still remains unclear \cite{Kantor87b,Kantor87c,Kownachi09}
whether the crumpling transition
(separating the flat and crumpled phases)
is critical 
\cite{Kantor86,Kantor87,Baig89,Ambjorn89,Renken96,Harnish96,Baig94,%
Bowick96b,Wheather93,Wheater96,David88,Doussal92,Espriu96} 
or belongs to a discontinuous one
with an appreciable latent heat \cite{Paczuski88,Kownacki02,Koibuchi04}.
%With the numerical simulations, however, it is not quite
%obvious to rule out a possibility
%of a weak-first-order transition \cite{Kantor87b,Kantor87c};
%see also Ref. \cite{Kownachi09}.

In this paper,
we investigate a discretized version of the polymerized
membrane
embedded in two dimensions
\cite{Kantor90,DiFrancesco94a,DiFrancesco94b,Cirillo96b};
details are overviewed afterward. %%%%explicated in Sec. \ref{section2}.
This model, the so-called discrete planar folding,
exhibits a series of crumpling transitions at $K \approx -0.3$ 
and $0.1$ \cite{DiFrancesco94b,Cirillo96b},
as the bending rigidity $K$ changes.
The latter transition exhibits a pronounced discontinuous
character, 
whereas the nature of the former transition remains unclear.  %%; details are overviewed in Sec. \ref{section2}.
In this paper, we utilized the transfer-matrix method 
\cite{DiFrancesco94b}
for the system sizes $L \le 14$.
We implemented a modified folding rule \cite{Nishiyama10},
Eq. (\ref{transfer_matrix}),
which enables us to impose the periodic-boundary condition.
Technically, the restoration of the translational symmetry
admits a substantial reduction of the transfer-matrix size.

%The rest of this paper is organized as follows.
%In Sec. \ref{section2},
%details of 
%the discrete planar folding are explained.
%In Sec. \ref{section3},
%we analyze the crumpling transition 
%in the $K<0$ side.
%The simulation method is explicated
%in the Appendix.
%In Sec. \ref{section4},
%we present the summary and discussions.

%\section{\label{section2}
%An overview of the discrete planar folding}
%In this section, we overview the discrete planar folding.

\begin{figure}
\includegraphics{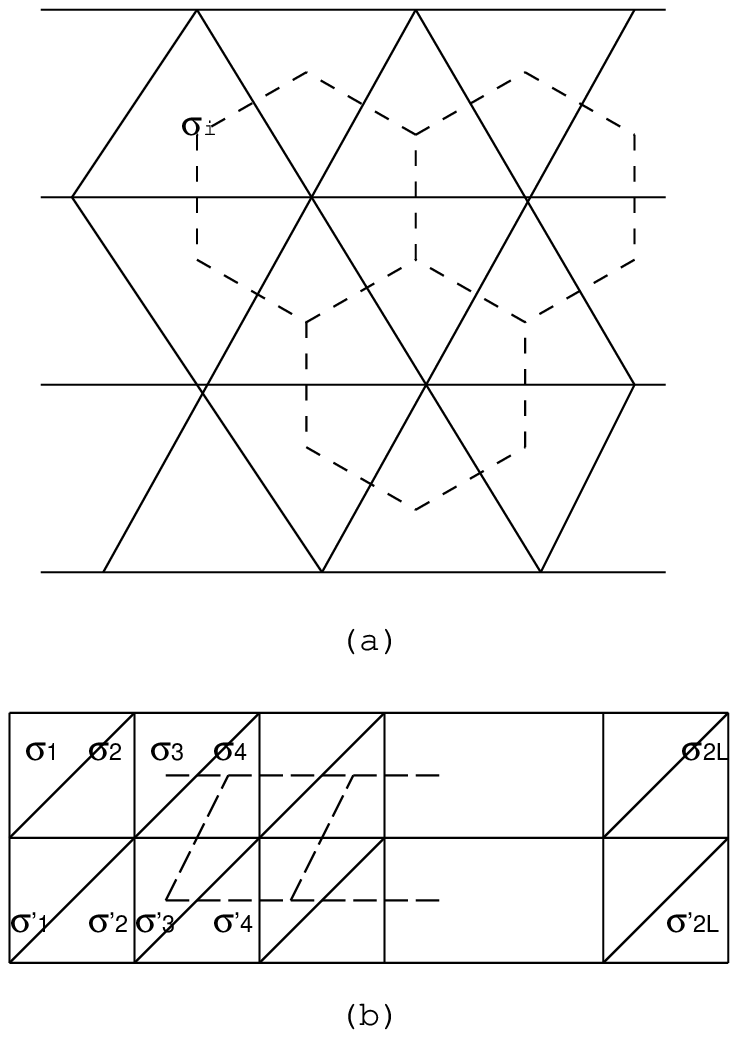}%
\caption{\label{figure1}
(a)
We consider a discrete folding of the triangular lattice.
The fold angle (with respect to the adjacent triangular plaquettes)
is discretized into either $\theta=0$ or $\pi$.
%In order to specify the fold angle, we place the Ising variables 
%$\{ \sigma_i \}$ on
%each triangle $i$
%rather than at each joint \cite{DiFrancesco94b}.
%Hence, hereafter, we consider a spin model on the
%dual (hexagonal) lattice.
(b)
A drawing of a transfer-matrix strip is shown.
%The row-to-row statistical weight yields the transfer-matrix element, 
%Eq. (\ref{transfer_matrix_original}).
%So far, the open-boundary condition has been imposed.
%Here, we restore the translational invariance by using a modified folding rule
%(\ref{transfer_matrix}).
}
\end{figure}

To begin with, we explain a basic feature of the discrete planar folding
\cite{DiFrancesco94b,Cirillo96b};
see Fig. \ref{figure1} (a).
We consider a sheet of the triangular lattice.
Along the edges, the sheet folds up.
The fold angle $\theta$ is either 
$\theta=0$ (complete fold) or $\pi$ (no fold).
The elastic energy at each edge is given by $K \cos \theta$ with the 
bending rigidity $K$.
%The transfer-matrix element of the planar folding 
%is given by Eq. (\ref{transfer_matrix_original}) \cite{DiFrancesco94b}.
%Here,
%the variables $\{ \sigma_i \}$ are 
%the Ising spins on the dual lattice;
%the dual transformation is explained in the Appendix.
%Apart from the prefactor of Kronecker's symbol,
%the transfer-matrix element (\ref{transfer_matrix_original})
%resembles to that of the ordinary Ising ferromagnet.
%The prefactor originates from the folding rule (constraint).
%The constraint is incompatible with the periodic-boundary condition,
%and so far, the open-boundary condition has been imposed
%\cite{DiFrancesco94b}.
%In this paper, 
%we implement the periodic-boundary condition
%through resorting to a modified folding rule \cite{Nishiyama10}, Eq. (\ref{transfer_matrix}).
%%The parameter $p$ in Eq. (  ) controls the modification;
%%namely, at $p=0$, the formula (  ) reduces to a conventional one.
%Technically, the restoration of the translational invariance
%renders a significant reduction of the
%transfer-matrix size.
%Taking the advantage, we treat system sizes up to $L=14$.
%
%
The thermodynamic property of the planar folding
has been studied extensively \cite{DiFrancesco94b,Cirillo96b}.
The transfer-matrix simulation for the system sizes $L \le 9$
\cite{DiFrancesco94b}
revealed a series of crumpling transitions at
$K \approx -0.3$ and $K = 0.11(1)$.
The behavior of the specific heat around $K \approx -0.3$
indicates that this transition would be a continuous one.
%%On the one hand, a pronounced discontinuous character is observed as to
%%the latter transition.
%%The nature of the former transition is less understood.
%In this paper, treating $L \le 14$,
%we analyze the singularity of the former transition in detail.
The cluster variation method (CVM)
of a single-hexagon-cluster approximation
\cite{Cirillo96b}
indicates that there occur crumpling transitions at
$K=-0.284$ and $K=0.1013$
of the continuous and discontinuous characters, respectively.

The crumpling transition $K \approx -0.3$
is closely related \cite{Bowick97}
to that of an extended folding 
\cite{Bowick95,Cirillo96,Bowick97}
at $K_3 \approx -0.8$.
(The extended folding, the so-called three-dimensional folding, 
has four possibilities,
$\cos \theta=\pm1,\pm1/3$,
 as to the joint angle $\theta$.)
That is, 
according to an argument based on a truncation of the
configuration space \cite{Bowick97},
the following (approximate) relations should hold;
\begin{eqnarray}
\label{relation1}
K  &=& K_3/3    \\
\label{relation2}
Q  &=& Q_3  .
\end{eqnarray}
Here, the variables $Q$ and $Q_3$ denote the latent heat 
for the planar- and three-dimensional-folding models, respectively.
A number of results, 
$(K_3,Q_3)=(-0.852,0)$ \cite{Bowick97}, 
$(-0.76(1),0.03(2))$ \cite{Nishiyama05}, and 
$(-0.76(10),0.05(5))$ \cite{Nishiyama10},
have been obtained via the CVM, density-matrix renormalization group, 
and exact-diagonalization analyses, respectively.
The nature of its transition at $K_3 \approx -0.8$ is not fully clarified,
because the three-dimensional folding is computationally demanding.
It is a purpose of this paper to shed light on
this longstanding issue from the viewpoint of the planar folding.
(It has to be mentioned that the planar folding has to a relevance to
a wide class of systems \cite{Moore04,Castelnovo05,Fendley07,DiFrancesco94a,Renken91}.)

%Last, we mention a number of related topics.
%First, the underlying physics of
%the planar folding is relevant to
%the Josephson-junction array
%\cite{Moore04}
%and
%the geometrically-constrained system
%\cite{Castelnovo05,Fendley07}.
%At $K=0$, in particular,
%the planar folding is exactly solvable
%\cite{DiFrancesco94a}.
%Second, 
%the polymerized membrane 
%embedded
%in two dimensions $d=2$ \cite{Renken91}
%may have a close relationship with the present study;
%the case of $d=3$ has been studied extensively,
%as mentioned in the Introduction.
%The $d=2$ membrane exhibits a pronounced 
%discontinuous character at the crumpling transition.
%(However, the underlying mechanism of this transition
%may not be quite relevant to that of the $K<0$ planar folding.)

%\appendix*
%\section{\label{appendix}
%Transfer-matrix formalism for the discrete planar folding}

For the sake of selfconsistency,
%In this Appendix,
we present the transfer-matrix formalism
for the discrete planar folding explicitly.
%%As mentioned in Sec. \ref{section3},
%A sheet of the triangular lattice
%[Fig. \ref{figure1} (a)]
%folds up
%along the edges.
%The fold angle $\theta$ is discretized into
%either $\theta=0$ (complete fold) or $\pi$ (no fold).
%The discretization leads to an Ising-spin representation
%of the discrete folding.
We place the Ising variables $\{ \sigma_i \}$
at each triangle $i$ (rather than each joint); see Fig. \ref{figure1} (a).
Hereafter, we consider the spin model on the dual (hexagonal) lattice.
The Ising-spin configuration specifies each joint angle between the adjacent triangles.
That is,
provided that the spins are (anti)parallel,
$\sigma_i \sigma_j=1$ $(-1)$,
for a pair of adjacent neighbors, $i$ and $j$, the joint angle is $\theta=\pi$ ($0$).
The spin configuration is 
subjected to a constraint (folding rule); the prefactor 
of the transfer-matrix element, Eq. (\ref{transfer_matrix_original}), enforces the constraint.
As a consequence, the discrete folding reduces to an Ising model
on the hexagonal lattice.
%(Because of the constraint mentioned above,
%the Monte Carlo method is not very efficient.)
In Fig. \ref{figure1} (b),
a drawing of
the transfer-matrix strip is presented.
The row-to-row statistical weight 
$T_{ \{\sigma_i \},\{\sigma'_i\} }$ yields the transfer-matrix element.
The transfer-matrix element for the strip length $L$ is given by \cite{DiFrancesco94b}
\begin{equation}
\label{transfer_matrix_original}
T_{  \{ \sigma'_i \} , \{ \sigma_i \} } =
 (  \prod_{i=1}^L
    \delta(\sigma_{2i}+\sigma_{2i+1}+\sigma_{2i+2}
          + \sigma'_{2i-1} + \sigma'_{2i}+\sigma'_{2i+1} 
\ mod \  3 ,0)
  )
  \exp(- \sum_{i=1}^L H_i(K)/T)
  ,
\end{equation}
with the local Hamiltonian
\begin{equation}
%H(K)=-0.5 K \sum_{i=1}^2L   (\sigma_i \sigma_{i+1}+ \sigma'_i\sigma'_{i+1})
%    -K \sum_{i=1}^L \sigma_{2i} \sigma'_{2i-1}
H_i(k)=-\frac{K}{2}
(
\sigma_{2i} \sigma_{2i+1}
+\sigma_{2i+1} \sigma_{2i+2}
+\sigma_{2i+2} \sigma'_{2i+1}
+\sigma'_{2i+1} \sigma'_{2i}
+\sigma'_{2i} \sigma'_{2i-1}
+\sigma'_{2i-1} \sigma_{2i}
)                          ,
\end{equation}
due to the bending-energy cost
for spins surrounding each hexagon $i$.
Here, the parameter $K$ denotes the bending rigidity,
and the expression $\delta(n,m)$ is Kronecker's symbol.
The periodic-boundary condition 
$\sigma_{L+i}=\sigma_{i}$ is imposed.
We set $T=1$, considering it as a unit of energy.

In practice, the above scheme does not work.
The folding rule is too restrictive to impose the periodic-boundary condition.
So far, the open-boundary condition has been implemented;
more specifically, the range of the running index $i$ in Eq. (\ref{transfer_matrix_original})
was set to $1 \le i \le L-1$ \cite{DiFrancesco94b}.
In this paper, following Ref. \cite{Nishiyama10},
we make a modification as to the constraint
(prefactor of Eq. (\ref{transfer_matrix_original})) to surmount the difficulty.
We replace the above expression with
\begin{equation}
\label{transfer_matrix}
T_{  \{ \sigma'_i \} , \{ \sigma_i \} } =
\frac{1}{L}\sum_{l=1}^L
 (  \prod_{i \ne l}
    \delta(\sigma_{2i}+\sigma_{2i+1}+\sigma_{2i+2}
          + \sigma'_{2i-1} + \sigma'_{2i}+\sigma'_{2i+1} 
\  mod \  3 ,0)
  )
  \exp(- \sum_{i \ne l} H_i(K) - H_l(K'))
.
\end{equation}
That is, the constraint is released at 
a defect hexagon $i=l$.
%%Moreover,
Additionally,
the local bending rigidity
at the defect is set to $K'$.
In order to improve the finite-size behavior,
we adjust $K'$ to
\begin{equation}
\label{defect_parameter}
K'=2K  .
\end{equation}
A justification is shown 
afterward.  %%) %%in Sec. \ref{section3_3}.)
%A single defect does not influence the thermodynamic (bulk)
%properties.
%Because the location of the defect distributes uniformly
%(via the summation $(1/L)\sum_{l=1}^L$),
%the translational invariance is maintained.
%%%%%%%%%%%%%%%%%%%The simulation results are presented in Sec. \ref{section3}.

%\section{\label{section3} Numerical results}
%In this section,
%%we present the simulation results.
%%We employed the transfer-matrix method (Appendix) 

Based on the transfer-matrix formalism
with a modified folding rule 
(\ref{transfer_matrix}),
we simulated the planar folding numerically.
The numerical diagonalization was performed within a subspace
specified by the wave number $k=0$
and the parity even;
here, we made use of the spin-inversion symmetry $\sigma_i \rightarrow -\sigma_i$.
%As mentioned in the Appendix,
%the defect parameter is adjusted to Eq. (\ref{defect_parameter});
%a justification of this choice
%is given in Sec. \ref{section3_3}.

%\subsection{Crumpling transition point}

\begin{figure}
\includegraphics{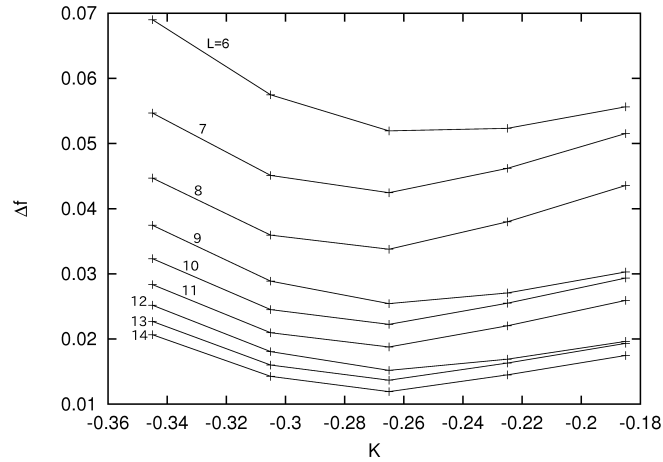}%
\caption{\label{figure2}
The free-energy gap (\ref{free_energy_gap})
is plotted for the bending rigidity $K$ and the
system sizes $6 \le L \le 14$.
%The data indicate that a crumpling transition
%(closure of $\Delta f$) occurs around $K \approx -0.27$; detailed analysis
%of this singularity is made in Fig. \ref{figure3}.
}
\end{figure}

In Fig. \ref{figure2},
we plot the free-energy gap
\begin{equation}
\label{free_energy_gap}
\Delta f =f_2-f_1            ,
\end{equation}
for the bending rigidity $K$ and various system sizes
$L=6,7,\dots,14$.
Here, the free energy per unit cell is 
given by $f_i=- \ln \Lambda_i/(2L)$
with the (sub)dominant eigenvalue $\Lambda_{1(2)}$
of the transfer matrix.
[Here, the unit cell stands for a triangle of the original lattice
rather than a hexagon of the dual lattice; 
see Fig. \ref{figure1}.]
From Fig. \ref{figure2},
we see a signature of a crumpling transition 
(closure of $\Delta f$) at 
$K \approx - 0.27$.
The location of the transition point
appears to be consistent with the preceding estimates \cite{DiFrancesco94b,Cirillo96b}. %% ;
%see Sec. \ref{section2}.

\begin{figure}
\includegraphics{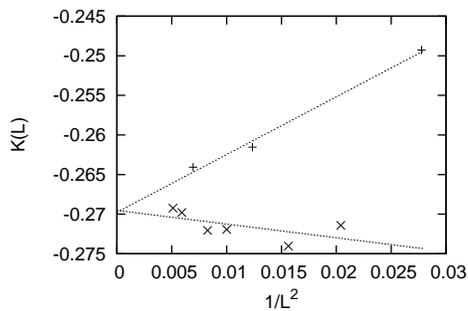}%
\caption{\label{figure3}
The transition point $K(L)$ (\ref{transition_point})
is plotted for $1/L^2$.
The linear least-squares fit for ($+$) $L=0$ and
($\times$) $1,2$ mod $3$ ($6 \le L \le 14$)
yields $K=-0.2697(12)$ and $-0.2695(14)$, respectively.
%%The symbols $+$ and $\times$ denotes the system sizes
%%of $L=0$ and $1,2$, respectively.
}
\end{figure}

In Fig. \ref{figure3},
the approximate transition point $K(L)$
is plotted for $1/L^2$ and 
$6 \le L \le 14$.
The approximate transition point
minimizes $\Delta f$;
namely  the relation
\begin{equation}
\label{transition_point}
\partial_K \Delta f |_{K=K(L)}=0
        ,
\end{equation}
holds.
%%%%%%%%%%%%%%%%%5 kufuu
The least-squares fit to 
a series of results for $L=6,9,12$
yields an estimate $K=-0.2697(12)$ in the thermodynamic limit $L\to \infty$.
Similarly, for $L=1,2$ mod $3$,
we obtain $K=-0.2695(14)$.
(An observation that the data $L=0$ and $1,2$ mod $3$ behave 
differently was noted in Ref. \cite{DiFrancesco94b}.)
The above independent results appear to be 
consistent with each other, validating the $1/L^2$-extrapolation scheme.
As a result,
we estimate the transition point as 
\begin{equation}
K=-0.270(2)
.
\end{equation}

%\subsection{Latent heat: Hamer's method}
%%In this section, we 

We then proceed to
estimate the amount of the latent heat 
with
Hamer's method
\cite{Hamer83}.
A basis of this method is as follows.
At the first-order transition point,
the low-lying spectrum of the transfer matrix
exhibits a level crossing, and the discontinuity
(sudden drop) of the slope reflects a
release of the latent heat.
However, the finite-size artifact
(level repulsion) smears out
the singularity.
According to Hamer \cite{Hamer83},
regarding the low-lying levels as nearly degenerate,
one can resort to the perturbation theory of the degenerated case,
and calculate the level splitting (discontinuity of slope) explicitly.
To be specific, we consider the matrix
\begin{equation}
\label{perturbation_matrix}
V=
\left(   
\begin{array}{cc}
V_{1 1}  &  V_{1 2}  \\
V_{2 1}  &  V_{2 2}
\end{array}
\right)
       ,
\end{equation}
with 
$V_{ij}=\langle i |  \partial_K T | j \rangle$
and the transfer matrix $T$.
The bases $| 1\rangle$
and $|2\rangle$ are the (nearly degenerate)
eigenvectors of $T$ with the eigenvalues $\Lambda_{1,2}$, respectively.
The states $\{ | i \rangle \}$ are normalized
so as to satisfy 
$\langle i |   T |  i \rangle =1$.
According to the perturbation theory,
the eigenvalues of Eq.
(\ref{perturbation_matrix})
yield the level-splitting slopes due to $K$.
Hence,
the latent heat
(per unit cell)
is given by a product of this discontinuity
and the coupling constant $K(L)$
\begin{equation}
\label{latent_heat}
Q (L) =       | K (L) |
\sqrt{  (V_{11}-V_{22})^2+4V_{12}V_{21} }   
          \frac{1}{2L}
,
\end{equation}
for the system size $L$.

\begin{figure}
\includegraphics{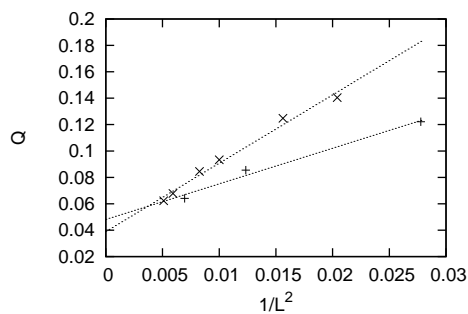}%
\caption{\label{figure4}
The latent heat $Q(L)$ (\ref{latent_heat})
is plotted for $1/L^2$. 
The linear least-squares fit for ($+$) $L=0$ and
($\times$) $1,2$ mod $3$ ($6 \le L \le 14$)
yields $Q=0.0482(59)$ and $0.0391(38)$, respectively.
%%The symbols are the same as those of Fig. \ref{figure3}.
}
\end{figure}

%%%%%%%%%%%%%%%%%%%%%%%%%%%%%%%
In Fig. \ref{figure4},
we plot the latent heat 
$Q$
(\ref{latent_heat})
for $1/L^2$ and $6 \le L \le 14$.
The least-squares fit 
for $L=6,9,12$ yields an estimate 
$Q=0.0482(59)$ in the
thermodynamic limit $L\to \infty$.
Similarly, for $L=1,2$ mod $3$,
we obtain $Q=0.0391(38)$.
Considering the deviation of these results as a possible systematic error,
we obtain
\begin{equation}
Q=0.043(10)
.
\end{equation}
The error margin covers 
both the statistical and systematic errors.

%%%%%%%%%%%%%%%%%%%%%%%%%
We consider the $1/L^2$-extrapolation scheme. 
%%%
The finite-size data are expected to converge rapidly (exponentially)
to the thermodynamic limit around the first-order transition point
for periodic boundary conditions,
%The finite-size 
%data converge rapidly (exponentially) to the thermodynamic limit around the first-order
%transition point, 
because the correlation length (typical length scale)
$\xi$ remains finite. Hence, the dominant finite-size corrections
in our case
should be described by $1/L^2$ (rather than $1/L$).
On the one hand, the curve of Fig. \ref{figure4}
appears to be concave down,
indicating an existence of a 
correction of O$(1/L)$.
However, this possibility (second-order phase transition) should be excluded:
In a preliminary stage, we made a finite-size-scaling analysis,
and arrived at a conclusion that the scaling theory does not apply;
the critical index $\nu$ estimated from the excitation gap
tends to diverge as $L\to\infty$.
Therefore, we set
the abscissa scale of Fig. \ref{figure4} to  $1/L^2$; 
actually,
the result of Fig. \ref{figure3}
demonstrates that 
the abscissa scale $1/L^2$
is sensible.
%Possibly, the convex character of Fig. \ref{figure4}
%should be attributed to a finite-size artifact which is not fully resolved yet.
%Even by the restoration of the translational invariance,
%a shaky character of the extrapolation curve,
%particularly, for the series of
%$L=1,2$ mod $3$, is not
%fully suppressed.

%%\subsection{\label{section3_3}
%%Simulation at $K'=0$}

\begin{figure}
\includegraphics{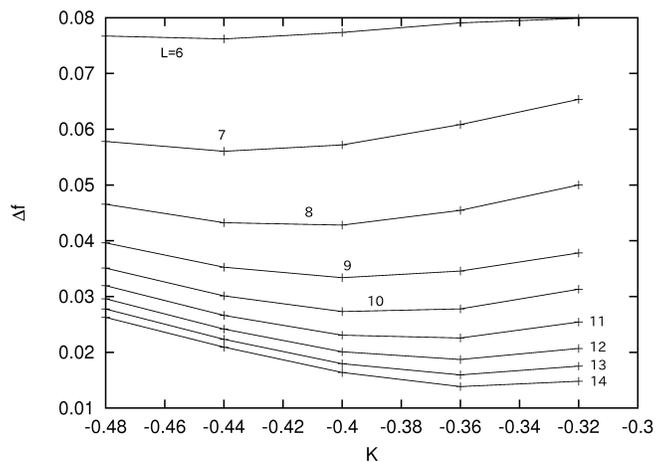}%
\caption{\label{figure5}
The free-energy gap (\ref{free_energy_gap}) is plotted for the bending rigidity $K$
and the system sizes $6 \le L \le 14$.
Tentatively, the defect parameter (\ref{transfer_matrix}) is set to $K'=0$.
%The data appear to be scatted as compared to those of Fig. \ref{figure2}.
%This result indicates that the finite-size behavior is influenced
%significantly by $K'$.
}
\end{figure}

As a comparison,
we provide a simulation result,
setting the defect parameter to $K'=0$
tentatively.
%The value $K'=0$ has an interpretation 
%such that
%a rupture
%(pair of open edges) distributes uniformly along the transfer-matrix strip.
%(This situation is an extension of the open-boundary condition,
%for which the rupture is static.)
%
In Fig. \ref{figure5},
we present the free-energy gap 
$\Delta f$ for the bending rigidity $K$;
the scale of $K$ is the same as that of Fig. 
\ref{figure2},
Clearly, the data of 
Fig. \ref{figure5} are less conclusive.
As a matter of fact,
the signatures of the crumpling transition 
strongly depend on the system size $L$.
This result indicates that the choice of the defect parameter 
$K'$ affects the finite-size behavior.
In the preliminary stage,
we survey a parameter space of $K'$,
and arrive at a conclusion that the above choice, 
Eq. (\ref{defect_parameter}), is an optimal one.
%Note that the parameter $K'$ is a
%byproduct of the modification of the folding rule, 
%Eq. (\ref{transfer_matrix}).
%Here, we make use of this redundancy
%so as to improve the finite-size behavior,
%aiming to take the thermodynamic limit reliably.

%%\section{\label{section4}
%%Summary and discussions}

In summary,
the crumpling transition 
of the discrete planar folding 
in
the $K<0$ regime was investigated
with the transfer-matrix method for $L \le 14$.
We adopted a modified-folding rule (\ref{transfer_matrix}),
which enables us to implement the periodic-boundary condition.
As a result,
we estimate the transition point and the latent heat
as 
$K=-0.270(2)$ and 
$Q=0.043(10)$, respectively.
%%
%% As mentioned in Sec. \ref{section2},
The planar- and three-dimensional-folding models are closely related; 
see Eqs. (\ref{relation1}) and (\ref{relation2}).
Making use of 
$K_3=-0.76(1)$ \cite{Nishiyama05}
and the present result $K=0.270(2)$,
we arrive at $K_3/K= 2.815(43) (\sim 3)$.
The relation (\ref{relation1}) appears to hold satisfactorily;
a slight deviation indicates that the truncation of the configuration space
is not exactly validated.
Encouraged by this result,
we estimate $Q_3=0.043(10)$ via Eq. (\ref{relation2}).
This result is consistent with 
$Q_3=0.03(2)$ \cite{Nishiyama05} and 
$Q_3=0.05(5)$ \cite{Nishiyama10},
indicating that the singularity belongs to a weak-first-order transition rather 
definitely.
Because a direct approach to the three-dimensional folding
is computationally demanding,
an indirect information from the planar folding would be valuable.
A further justification of the configuration-space truncation 
would be desirable to confirm this claim.
This problem will be addressed in the future study.

% If in two-column mode, this environment will change to single-column
% format so that long equations can be displayed. Use
%\begin{widetext}
%\end{widetext}

% figures should be put into the text as floats.
% Use the graphics or graphicx packages (distributed with LaTeX2e)
% and the \includegraphics macro defined in those packages.
% See the LaTeX Graphics Companion by Michel Goosens, Sebastian Rahtz,
% entire page. There is no need to do explicit centering.

\end{document}